# Improved spot formation for flexible multi-mode fiber endoscope using partial reflector

**Ruo Yu Gu[1,*], Elaine Chou[1], Cory Rewcastle[2], Ofer Levi[2], and Joseph M. Kahn[1]**

[1]*E. L. Ginzton Laboratory and Department of Electrical Engineering,*
*Stanford University, Stanford, CA 94305*
[2]*Institute of Biomaterials and Biomedical Engineering, University of Toronto, 164 College Street,*
*Toronto, Ontario M5S 3G9, Canada*
*\*ruoyugu@stanford.edu*

**Abstract:** Multi-mode fiber (MMF) endoscopes are a new type of endoscope that use only a single optical fiber to transmit images, achieving much higher spatial resolution for a given diameter than commercial fiber bundle endoscopes. However, a drawback of MMF endoscopes is that imaging quality degrades substantially as the MMF is perturbed. We propose a method for improving spot formation at the distal end of a perturbed MMF, and thus improving imaging quality, by attaching a partial reflector to the distal end of the MMF. We experimentally find that the perturbation of the light reflected from the partial reflector is highly correlated with the perturbation of the light transmitted through the MMF. We demonstrate a simple method exploiting this correlation that enables formation of spots at the distal fiber end with quality 10-15% higher than if the perturbation of the MMF is ignored. In the future, more advanced algorithms exploiting the correlation may enable further improvements in spot formation and imaging through perturbed fibers.

## References and links

1. B. A. Flusberg, E. D. Cocker, W. Piyawattanametha, J. C. Jung, E. L. M. Cheung, and M. J. Schnitzer, "Fiber-optic fluorescence imaging," Nat. Methods **2**, 941–950 (2005).
2. E. R. Andresen, S. Sivankutty, G. Bouwmans, L. Gallais, S. Monneret, and H. Rigneault, "Measurement and compensation of residual group delay in a multi-core fiber for lensless endoscopy," J. Opt. Soc. Am. B **32**, 1221–1228 (2015).
3. Y. Kim, S. C. Warren, J. M. Stone, J. C. Knight, M. A. A. Neil, C. Paterson, C. W. Dunsby, and P. M. W. French, "Adaptive Multiphoton Endomicroscope Incorporating a Polarization-Maintaining Multicore Optical Fibre," IEEE J. Sel. Top. Quantum Electron. **22**, 1–8 (2016).
4. S. C. Warren, Y. Kim, J. M. Stone, C. Mitchell, J. C. Knight, M. A. A. Neil, C. Paterson, P. M. W. French, and C. Dunsby, "Adaptive multiphoton endomicroscopy through a dynamically deformed multicore optical fiber using proximal detection," Opt. Express **24**, 21474–21484 (2016).
5. D. B. Conkey, N. Stasio, E. E. Morales-Delgado, M. Romito, C. Moser, and D. Psaltis, "Lensless two-photon imaging through a multicore fiber with coherence-gated digital phase conjugation," J. Biomed. Opt. **21**, 45002 (2016).
6. N. Stasio, D. B. Conkey, C. Moser, and D. Psaltis, "Light control in a multicore fiber using the memory effect," Opt. Express **23**, 30532–30544 (2015).
7. S. M. Kolenderska, O. Katz, M. Fink, and S. Gigan, "Scanning-free imaging through a single fiber by random spatio-spectral encoding," Opt. Lett. **40**, 534–537 (2015).
8. R. N. Mahalati, R. Y. Gu, and J. M. Kahn, "Resolution limits for imaging through multi-mode fiber," Opt. Express **21**, 2375–2384 (2013).
9. M. Plöschner and T. Čižmár, "Compact multimode fiber beam-shaping system based on GPU accelerated digital holography," Opt. Lett. **40**, 197–200 (2015).
10. A. M. Caravaca-Aguirre, E. Niv, and R. Piestun, "High-Speed Phase Modulation for Multimode Fiber Endoscope," Imaging Appl. Opt. (2014).
11. M. Plöschner, B. Straka, K. Dholakia, and T. Čižmár, "GPU accelerated toolbox for real-time beam-shaping in multimode fibres," Opt. Express **22**, 2933–2947 (2014).
12. D. Loterie, S. Farahi, I. Papadopoulos, A. Goy, D. Psaltis, and C. Moser, "Digital confocal microscopy through a multimode fiber," Opt. Express **23**, 23845–23858 (2015).

13. D. Loterie, D. Psaltis, and C. Moser, "Confocal microscopy via multimode fibers: fluorescence bandwidth," SPIE BIOS **9717**, 97171C (2016).
14. E. E. Morales-Delgado, D. Psaltis, and C. Moser, "Focusing and scanning of femtosecond pulses through a multimode fiber: applications in two-photon imaging and polymerization," presented at the Australian Conference on Optical Fibre Technology, Sydney, Australia, 6–8 Sep. 2016.
15. E. E. Morales-Delgado, D. Psaltis, and C. Moser, "Two-photon imaging through a multimode fiber," Opt. Express **23**, 32158–32170 (2015).
16. E. E. Morales Delgado, D. Psaltis, and C. Moser, "Two-photon excitation endoscopy through a multimode optical fiber," SPIE BIOS **9717**, 97171E (2016).
17. S. Sivankutty, E. R. Andresen, R. Cossart, G. Bouwmans, S. Monneret, and H. Rigneault, "Ultra-thin rigid endoscope: two-photon imaging through a graded-index multi-mode fiber," Opt. Express **24**, 825–841 (2016).
18. S. Ohayon, A. M. Caravaca-Aguirre, R. Piestun, and J. J. DiCarlo, "Deep brain fluorescence imaging with minimally invasive ultra-thin optical fibers," arXiv Prepr. arXiv1703.07633 (2017).
19. Y. Xu, Y. Lu, Z. Zhu, J. Shi, C. Guan, and L. Yuan, "Multimode fiber focusing lens based on plasmonic structures," SPIE/COS Photonics Asia, **10019**, 100191C (2016).
20. S. Bianchi, V. P. Rajamanickam, L. Ferrara, E. Di Fabrizio, C. Liberale, and R. Di Leonardo, "Focusing and imaging with increased numerical apertures through multimode fibers with micro-fabricated optics," Opt. Lett. **38**, 4935–4938 (2013).
21. Y. Choi, C. Yoon, M. Kim, W. Choi, and W. Choi, "Optical imaging with the use of a scattering lens," IEEE J. Sel. Top. Quantum Electron. **20**, 61–73 (2014).
22. S. Rosen, D. Gilboa, O. Katz, and Y. Silberberg, "Focusing and Scanning through Flexible Multimode Fibers without Access to the Distal End," arXiv Prepr. arXiv1506.08586 (2015).
23. T. Čižmár and K. Dholakia, "Optical manipulation, beam-shaping and scanner-free bright-field and dark-field imaging via multimode optical fibre," SPIE Photonics West, **8637**, 86370G (2013).
24. D. Loterie, D. Psaltis, and C. Moser, "Bend translation in multimode fiber imaging," Opt. Express **25**, 6263 (2017).
25. W. Xiong, P. Ambichl, Y. Bromberg, B. Redding, S. Rotter, and H. Cao, "Principal modes in multimode fibers: exploring the crossover from weak to strong mode coupling," Opt. Express **25**, 2709–2724 (2017).
26. A. M. Caravaca-Aguirre and R. Piestun, "Single multimode fiber endoscope," Opt. Express **25**, 1656–1665 (2017).
27. R. Y. Gu, R. N. Mahalati, and J. M. Kahn, "Design of flexible multi-mode fiber endoscope," Opt. Express **23**, 26905–26918 (2015).
28. D. Marcuse, *Theory of Dielectric Optical Waveguides*, 2nd ed. (Academic Press Inc., 1991).
29. R. N. Mahalati, D. Askarov, J. P. Wilde, and J. M. Kahn, "Adaptive control of input field to achieve desired output intensity profile in multimode fiber with random mode coupling," Opt. Express **20**, 14321–14337 (2012).
30. R. Y. Gu, R. N. Mahalati, and J. M. Kahn, "Noise-reduction algorithms for optimization-based imaging through multi-mode fiber," Opt. Express **22**, 15118–15132 (2014).
31. O. Geschke, H. Klank, and P. Telleman, eds., *Microsystem Engineering of Lab-on-a-Chip Devices* (John Wiley & Sons, 2004).
32. R. Horisaki, R. Takagi, and J. Tanida, "Learning-based imaging through scattering media," Opt. Express **24**, 13738–13743 (2016).
33. M. Lyu, H. Wang, G. Li, and G. Situ, "Exploit imaging through opaque wall via deep learning," arXiv Prepr. arXiv1708.07881 (2017).

## 1. Introduction

Endoscopes are medical instruments that transmit images through long, thin optical conduits. Small diameter and high resolution are important attributes of endoscopes, particularly when imaging deep tissue regions inaccessible to other imaging devices. Typical commercial endoscopes are constructed using a bundle of single-mode fibers (SMFs), and have diameters as small as about 0.5 mm, achieving a spatial resolution of about one thousand pixels [1]. By contrast, single-fiber endoscopes are a new type of endoscope that use only a single optical fiber to transmit images; typically this single optical fiber is a multi-mode fiber (MMF), though some methods use a multi-core fiber (MCF) [2–6] or a SMF [7]. Single-fiber endoscopes offer a better trade-off between diameter and spatial resolution than fiber bundle endoscopes: a single-fiber endoscope with 0.5 mm-diameter using a typical MMF would resolve more than half a million pixels [8]. Despite these advantages, endoscopes using a single MMF have weaknesses, including long imaging times, small field of view, and the requirement that the MMF be held rigid.

Research addressing all of these weaknesses has progressed at a rapid pace. To address long imaging times, new techniques have been developed to increase imaging speed [9–11] and signal-to-noise ratio (SNR) [12,13], in particular, two-photon excitation imaging [14–17]; and improvements in key components, especially spatial-light modulators, have substantially improved both speed and accuracy in recent years. In fact, both speed and SNR have improved to the point that real-time *in vivo* biological imaging through a single rigid fiber has been demonstrated [18]. In parallel, the field of view limitation is being addressed by the design of small passive optical components to be placed the distal end of the optical fiber [19–21]. Finally, many methods have been proposed to address the rigidity requirement to enable imaging through a flexible fiber. Among the most promising approaches are: imaging through a graded-index MMF by using multi-photon excitation endoscopy, exploiting the correlation properties of graded-index MMFs to create spots in known locations [22]; imaging through a SMF using broadband spectra by encoding spatial information in different wavelengths [7]; imaging through a flexible MCF by encoding image information in the different cores of a MCF and compensating for the change in relative phases [4]; and compensation for mode coupling in MMF using a perturbation-invariant basis by monitoring the physical configuration of the MMF [23]. Recent research has also found that mode coupling through a MMF is surprisingly weak for small perturbations [24–26]. Each method has strengths and weaknesses; for example, spatio-spectral encoding requires that the imaged object have known spectral characteristics [7]; MCFs are limited to approximately 100 cores in order to remain thin [4]; the method in [23] requires the MMF's physical configuration to be known [23].

In this paper, we extend and experimentally demonstrate a method first described in [27] for calibrating a flexible MMF by placing a structured reflector on the distal end of the MMF, which makes it possible to measure information on the mode coupling in the MMF based on measurement of light reflected from the partial reflector. A diagram showing how such a flexible endoscope would work before and after insertion into a sample to be imaged is shown in Fig. 1. The apparatus in Fig. 1 reflects two modifications as compared to the apparatus described in [27]. First, a structured reflector is not necessary, and a simple partial reflector suffices. Second, the imaging modality is fluorescence imaging instead of reflective imaging; because the light reflected from the partial reflector and fluorescence from the sample are at different wavelengths, the SNR is increased and a shutter, as described in [27], is unnecessary. We show in this paper that the perturbation of the modes reflected from a partial reflector at the distal end of the MMF is strongly correlated to the perturbation of the modes transmitted forward through the MMF, as predicted theoretically. This makes it possible to improve the quality of spots that can be formed through a perturbed MMF. The strong correlation between the perturbation of the transmitted and reflected light signals further suggests that a near one-to-one mapping between the two may be possible, which would allow for accurate imaging through a flexible MMF.

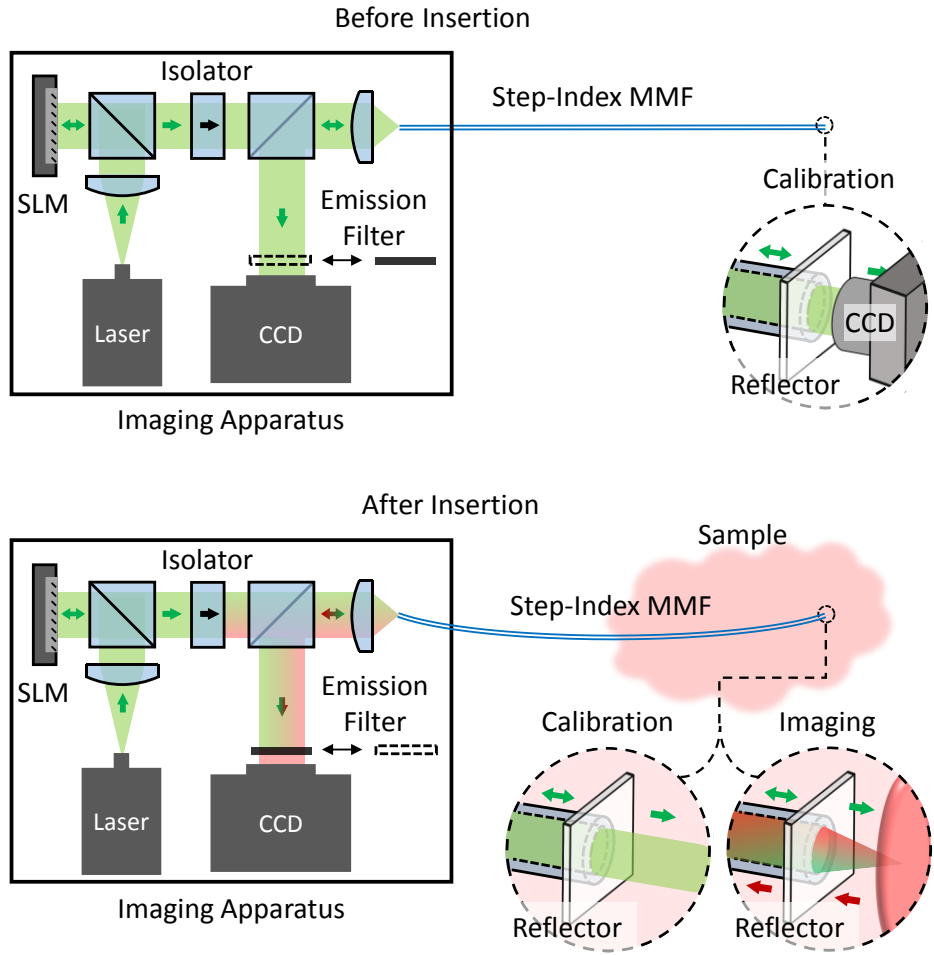

Fig. 1. Schematic showing the concept for a flexible MMF endoscope. The long-pass emission filter shown in front of the CCD camera can be moved in or out of the beam path. In a calibration performed before the MMF is inserted into the sample, the transmission of light through the MMF is measured by a camera at the distal end of the MMF, and the reflection of light from the partial reflector at the distal end of the MMF is measured by a camera at the proximal end of the MMF (without the emission filter). After this calibration, the MMF is inserted into the sample. Whenever the MMF is perturbed, the reflected light from the partial reflector is re-measured (without the emission filter), and this calibration allows imaging through the perturbed MMF (with the emission filter). The imaging modality is fluorescence imaging, which means the light emitted from the sample and collected by the MMF is at a longer wavelength than the light reflected from the partial reflector. Since light at the two wavelengths can be separated, the SNR is increased and a shutter at the distal end of the MMF is not needed.

The remainder of this paper is as follows. In Section 2, we briefly describe the theory describing the propagation of light in a MMF that we use throughout the paper. In Section 3, we describe our experimental apparatus and our procedure for coherent measurement of the wavefront of the light. In Section 4, we describe our method to compensate for the perturbation of the MMF using reflection measurements. In Section 5, we present our experimental results, and in Section 6, we discuss the results. We conclude in Section 7.

## 2. Mathematical background for light propagation inside MMF

Light propagating through a MMF can be concisely described in terms of the electric and magnetic fields of a set of orthogonal guided modes. For convenience we define the "proximal" and "distal" ends of the MMF, a "forward" direction (from proximal to distal) along the length of the MMF, and the opposite direction as "backward," all as shown in Fig. 2. At every point along the MMF we can then orient a local right-handed Cartesian coordinate system $(x, y, z)$ whose $+z$ axis is a tangent to the central axis of the MMF and points in the forward direction.

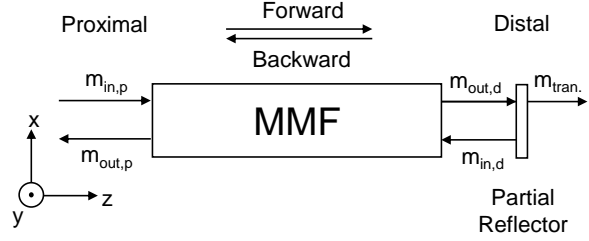

Fig. 2. Schematic of MMF, showing the Cartesian coordinate axes, proximal and distal ends of the MMF, forward and backward directions, and the labeling of the mode vectors. A partial reflector is attached directly to the distal end of the MMF (with no air gap).

At each $(x, y, z)$ location, the MMF supports so-called local normal modes [28], which are mathematically equivalent to the ideal TE, TM, HE and EH modes of a straight MMF with the same coordinate system. The precise $\vec{E}(x, y, z, t)$ and $\vec{H}(x, y, z, t)$ fields of the forward-propagating modes are described in [28], and equivalently using phasor notation as complex $\vec{E}(x, y, z)$ and $\vec{H}(x, y, z)$. There are an equal number of backward-propagating modes, which have fields $\vec{E}^*(x, y, z)$ and $-\vec{H}^*(x, y, z)$, where the asterisk represents the complex conjugate. Each mode is normalized to have unit power. We keep track of propagating modes only and ignore radiation modes.

Let $N$ be the number of forward-propagating modes in the MMF, which is also equal to the number of backward-propagating modes. The amplitude and phase of a mode at one plane in the MMF is described by a single complex coefficient; and the amplitude, phase and polarization of the electric and magnetic fields at that plane inside the MMF are thus represented by a single complex $2N \times 1$ vector **m**. We can make two further simplifications. First, for an endoscopic application we are only concerned with the modes at the proximal and distal facets of the MMF, just outside the fiber facet; for MMFs with low numerical aperture (NA), at these facets it is convenient to use the basis of linearly polarized (LP) modes instead of the more complicated basis of local normal modes [28]. Second, since forward and backward propagating modes do not couple to each other inside the MMF, it is convenient to separate **m** at the distal plane into $N \times 1$ vectors $\mathbf{m}_{in,d}$ and $\mathbf{m}_{out,d}$, and **m** at the proximal plane into $N \times 1$ vectors $\mathbf{m}_{in,p}$ and $\mathbf{m}_{out,p}$, as shown in Fig. 2. We order the coefficients in each of these vectors as follows: LP modes in the x-polarization are ordered by decreasing propagation constant, followed by LP modes in the y-polarization ordered by decreasing propagation constant.

The mathematical relation between the mode coefficients at the proximal and distal ends of the MMF can then be completely described in matrix notation using matrices **T** and **R**:

$$\begin{aligned} \mathbf{m}_{out,d} &= \mathbf{T}\mathbf{m}_{in,p} \\ \mathbf{m}_{out,p} &= \mathbf{R}\mathbf{m}_{in,d} \end{aligned}, \quad (1)$$

where **T** is by definition the transmission matrix of the MMF and **R** is by definition the reflection matrix of the MMF; both change if the MMF is physically perturbed. Eq. (1) fully describes the light at both ends of the MMF. For example, the (scaled) irradiance $I$ of the forward-propagating light at the distal end as it would appear on a properly calibrated CCD placed at the distal plane can be computed as:

$$I(x,y) = \left|\vec{\mathbf{E}}_{out,d}(x,y)\cdot\vec{\mathbf{n}}\right|^2 = \left|\left(\sum_{k=1}^{N}(m_{out,d})_k \vec{\mathbf{E}}_k(x,y)\right)\cdot\vec{\mathbf{n}}\right|^2 , \qquad (2)$$

where $\vec{\mathbf{E}}_k(x,y)$ is the electric field of LP mode $k$ and $\vec{\mathbf{n}}$ is the unit normal vector to the CCD surface.

### 3. Endoscope apparatus and coherent measurement procedures

The apparatus for measuring perturbation of the endoscope is shown in Fig. 3. For convenience, the apparatus can be divided into three sections: laser source section, endoscope section, and camera section. All lenses and non-polarizing beamsplitters (NPBS) are antireflection coated at visible wavelengths.

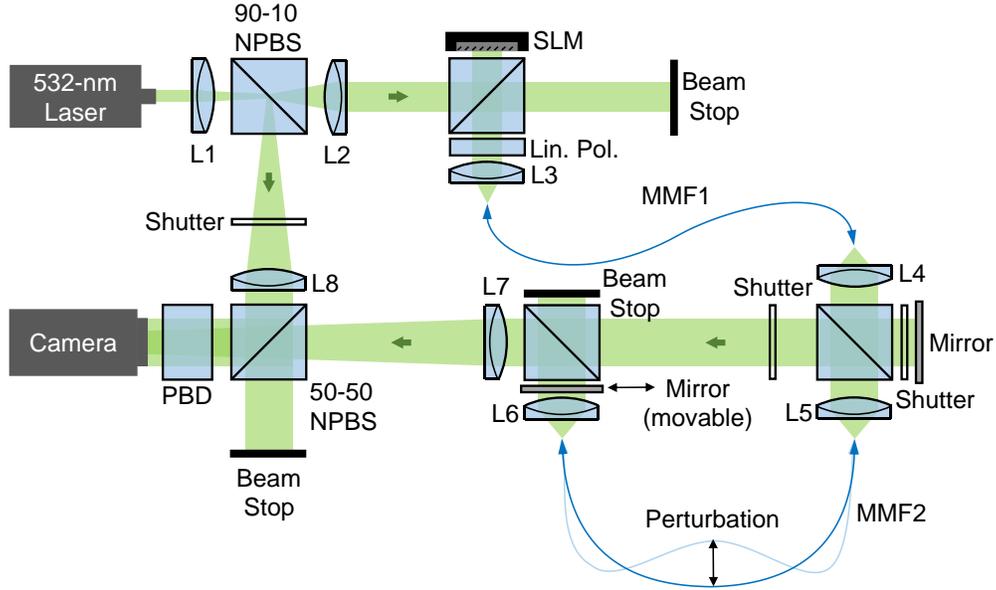

Fig. 3. Diagram of experimental setup for imaging through a perturbed multi-mode fiber (MMF2). The laser source sections, endoscope section and camera sections are shown. All three shutters are shown open. All beamsplitters are 50-50 non-polarizing beamsplitters except the labeled 90-10 non-polarizing beamsplitter.

The laser source section consists of a laser (single-longitudinal-mode, 200 mW, 532-nm, vertical polarization of beam), two achromatic doublet lenses (L1: $f$ = 19 mm, NA = 0.3 and L2: $f$ = 200 mm, NA = 0.06) for beam expansion, a 90-10 NPBS (90% transmitted, 10% reflected power), a 50-50 NPBS, a linear polarizer, a phase-only nematic liquid crystal-on-silicon reflective SLM (Meadowlark HSP256-0532, 256×256 pixels, each 24×24 μm² in size, with phases controllable from 0 to $2\pi$ with 8-bit resolution), an aspheric lens (L3: $f$ = 4.6 mm, NA = 0.53) to couple light into a fiber, and a MMF labeled MMF1 (Thorlabs custom order, 1 m long, 25 μm core diameter, 0.1 NA, FC/APC connector at proximal end, FC/PC connector at distal end). This arrangement allows a complete basis of LP modes to be excited at the distal end of MMF1 by setting different patterns onto the SLM and rotating the linear polarizer between configurations of −45°, 0° and 45° with respect to the polarization of the beam. Essentially, MMF1 acts as a mode filter. Also, part of the laser beam is split off by a NPBS to act as a reference beam for the camera section.

The endoscope section is a MMF (Thorlabs custom order, 1 m long, 25 μm diameter, 0.1 NA, FC/PC both ends) labeled MMF2, cut from the same fiber as MMF1 and re-connectorized; this ensures that both MMFs support precisely the same propagating modes. MMF2 is secured to the optical table, and two matching microscope objective lenses (L4 and L5: 20× magnification, 1.2 mm working distance, NA = 0.4) couple light from MMF1 to MMF2. A 50-50 NPBS between the objective lenses sends both the light exiting distally from MMF1 and the light exiting proximally from MMF2 to the camera section. At the distal end of MMF2, an aspheric lens (L6: $f$ = 10.0 mm, NA = 0.55) is used to collimate light. A 99% reflectivity silver reflector at the focal plane of the aspheric lens can be used to reflect light back into MMF2 with high coupling efficiency, or moved out of the way to send the collimated light to the camera section. This moveable reflector acts as the partial reflector; in a true endoscopic scenario a partial reflector would instead be attached directly to MMF2. In this way, light described by $\mathbf{m}_{in,p}$, $\mathbf{m}_{out,p}$ and $\mathbf{m}_{out,d}$ (see Fig. 2) can all be independently directed towards the camera section. During measurement of the MMF, only one of these beams, designated as the signal beam, is sent to the camera section at a time; the rest are blocked by shutters. By manually perturbing MMF2, the mode coupling of the MMF can be experimentally perturbed.

The camera section is comprised of a camera (QImaging Retiga 4000R, 2048×2048 pixels, each 7.4×7.4 μm$^2$ in size, with 10-bit resolution), a polarizing beam displacer (PBD) (2.7 mm displacement), an achromatic doublet lens to image the signal beam (L7: $f$ = 300 mm, NA = 0.04), an achromatic doublet lens to collimate the reference beam (L8: $f$ = 75 mm, NA = 0.09), and a shutter in the path of the reference beam. This camera setup allows coherent measurement of both polarizations of any of the three beams directed into the camera from the endoscope section.

From Eq. (1), full characterization of the transmission **T** and reflection **R** matrices of MMF2 requires measurements of $\mathbf{m}_{in,p}$, $\mathbf{m}_{out,p}$ and $\mathbf{m}_{out,d}$ of MMF2; this is accomplished as follows. First, a set of $M$ SLM patterns are computed that will excite as closely as possible an orthonormal set of modes at the distal end of MMF1. Such a computed set of $M$ patterns achieves a higher SNR than a random set of patterns. $M$ is at minimum the number of modes MMF2 supports in one polarization, $N/2$, but for robustness should be higher by a factor of two to three. When modulating the SLM, 6×6 pixel binning is used to approximate amplitude modulation using phase-only modulation [29]. For reasons that will be explained shortly, in all $M$ patterns, 15% of the SLM is set aside and set to amplitude 0. Pattern 1 is then placed onto the SLM, the linear polarizer is rotated to −45°, and $\mathbf{\vec{E}}_{in,p}(x, y)$ is incident upon the camera section while shutters block all other beams. $\mathbf{\vec{E}}_{in,p}(x, y)$ is measured coherently (i.e., measurement of amplitude and phase) by obtaining four measurements: the irradiance of the reference beam, the irradiance of the signal beam, the irradiance of the coherent interference between the signal and reference beams, and the irradiance of the coherent interference between the signal beam, shifted by a phase of π/2 radians with respect to the reference beam. This procedure is repeated for SLM pattern 2, etc., up until pattern $M$. However, since each successive coherent measurement is separated in time, each $\mathbf{\vec{E}}_{in,p}(x, y)$ measurement is not in phase with the others due to jitter between the phases of the reference and signal beams. This is remedied by using the 15% of the SLM set aside earlier to simultaneously excite a second reference beam that travels through the MMF along with the signal beam (thus this reference stays in phase with the signal beam), and taking a set of measurements of the interference between this reference and the signal beam for every $\mathbf{\vec{E}}_{in,p}(x, y)$. This ensures that all

$\mathbf{\tilde{E}}_{in,p}(x, y)$ are in phase. After that, the linear polarizer is rotated to 45° and the procedure is repeated for another $M$ measurements, so that complete bases of $x$- and $y$-polarized LP modes are independently excited. Since the coherent measurements of these two sets of modes are separated in time, another set of $M$ measurements taken at 0° rotation of the linear polarizer is used to match the phases of the 45° and −45° measurements. All these $\mathbf{\tilde{E}}_{in,p}(x, y)$ are then adjusted for $x$ and $y$ position, $x$ and $y$ tilt, and focus, by centering the average of all coherent measurements in both the spatial and spatial frequency domains. Once properly centered, these measurements are decomposed in an orthonormal basis of modes to obtain $\mathbf{m}_{in,p}$. This orthonormal basis is the LP mode basis for an ideal step-index MMF having the nominal core radius and NA of MMF2, slightly modified to more accurately account for spherical aberration, defocus of the lenses, and small deviations from nominal core radius and NA. Finally, this process is repeated to measure $\mathbf{m}_{out,p}$ and $\mathbf{m}_{out,d}$ for full characterization of the MMF. This process is repeated for every perturbation of MMF2.

## 4. Perturbation compensation using partial reflector

In this section we describe the theory behind compensating for perturbation of the MMF, and hence a perturbation of $\mathbf{T}$, by taking a measurement of the perturbation of $\mathbf{R}$. The quality of an image obtained through a MMF corresponds directly to how accurately any desired pattern can be excited at the distal end of the MMF [30]; for simplicity we will assume this pattern is a spot, and that the spot forms the point-spread function of the imaging system. The quality of the spot that can be formed in turn depends directly on how accurately $\mathbf{T}$ is known, since to form a spot, the transmission matrix must be inverted as shown in Eq. (3):

$$\begin{aligned}\mathbf{m}_{out,d} &= \mathbf{T}\mathbf{m}_{in,p} \\ \mathbf{m}_{out,d} &= \mathbf{T}(\mathbf{\tilde{T}}^{-1}\mathbf{\tilde{m}}_{out,d})\end{aligned} \quad , \quad (3)$$

where $\mathbf{\tilde{m}}_{out,d}$ are the desired mode coefficients at the distal end that correspond to a spot, and $\mathbf{\tilde{T}}$ is a noisy measurement or approximation of $\mathbf{T}$. The closer $\mathbf{T}\mathbf{\tilde{T}}^{-1}$ is to the identity matrix, the better the spot will be; if it is equal to the identity matrix, the smallest spot that can be supported by the MMF will be formed. For a rigid MMF, $\mathbf{T}$ can be measured precisely before imaging and does not change. We note that during imaging it is necessary to form spots through the partial reflector, as shown in Fig. 2, but no extra step is required because its transmission matrix is the (scaled) identity matrix.

Since $\mathbf{T}$ only depends on the current physical state of the MMF, imaging through a flexible MMF using Eq. (3) is possible if $\mathbf{T}$ is known for all possible perturbations within some allowed range of motion. To model this, let a MMF have a (known) initial and an (unknown) perturbed configuration, with all displacements between the two configurations be considered the perturbation. A partial reflector is placed at the distal end of the MMF, so that the reflection matrix $\mathbf{R}$ can be measured. Working from Eq. (1), the theoretical relationships between $\mathbf{T}$, $\mathbf{R}$, and the perturbed matrices are as follows:

$$\begin{aligned}\mathbf{m}_{out,d} &= \mathbf{TCm}_{in,p} & \mathbf{m}'_{out,d} &= \mathbf{P_t TCm}_{in,p} \\ \mathbf{m}_{out,p} &= \mathbf{ROm}_{out,d} & \mathbf{m}'_{out,p} &= \mathbf{P_r ROm}'_{out,d}\end{aligned} \quad , \quad (4)$$

where $\mathbf{m}'$ denotes the perturbed mode coefficients, the mode coupling into the MMF is described by a mode-coupling matrix $\mathbf{C}$, the mode coupling back into the MMF due to the partial reflector is described by a matrix $\mathbf{O}$, and $\mathbf{P_t}$ and $\mathbf{P_r}$ are perturbation matrices. There is no loss of generality in assuming that $\mathbf{T}$ is perturbed into $\mathbf{P_t T}$ (i.e., $\mathbf{T}$ multiplied by some perturbation matrix $\mathbf{P_t}$) and $\mathbf{R}$ into $\mathbf{P_r R}$ (i.e., $\mathbf{R}$ multiplied by some perturbation matrix $\mathbf{P_r}$) as long as both **T** and **R** are full-rank. We assume that the initial configuration of the MMF can be measured accurately, but that after the MMF is perturbed there is no access to the distal end; thus $\mathbf{m}_{in,p}$, $\mathbf{m}_{out,p}$ and $\mathbf{m}_{out,d}$ can all be measured before the MMF is perturbed; but only $\mathbf{m}'_{out,p}$ can be measured while $\mathbf{m}'_{out,d}$ cannot after the fiber is perturbed. The problem then becomes how to calculate $\mathbf{P_t}$ from Eq. (4).

If we further take into account reciprocity (so that $\mathbf{R} = \mathbf{T}^T$) [27], the equations become:

$$\mathbf{m}_{out,d} = \mathbf{TCm}_{in,p} \qquad \mathbf{m}'_{out,d} = \mathbf{P_t TCm}_{in,p}$$
$$\mathbf{m}_{out,p} = \mathbf{T}^T \mathbf{Om}_{out,d} \qquad \mathbf{m}'_{out,p} = (\mathbf{P_t T})^T \mathbf{Om}'_{out,d} \qquad (5)$$

From Eq. (5), we described a method to calculate $\mathbf{P_t}$ in [27]; however, that solution made the assumption that coupling into the MMF was perfect (i.e., $\mathbf{C} = \mathbf{I}$), with no mode coupling or mode-dependent loss. Though it is still possible to find a unique solution even with a non-ideal **C**, in a real system Eq. (5) is further complicated by experimental error and mode-dependent loss of **T**, violating one of the assumptions we made in [27]. This makes exact, closed-form solutions of Eq. (5) prone to error and overfitting. Instead, by matching Eq. (4) and Eq. (5) we can obtain the result:

$$\mathbf{P_t} = \mathbf{T}^{-1} \mathbf{P_r}^T \mathbf{T} \quad . \qquad (6)$$

This suggests that when $\mathbf{P_r}$ and **T** are close to diagonal, $\mathbf{P_r}^T$ should be similar to $\mathbf{P_t}$ and thus can be used as an approximation to $\mathbf{P_t}$. We would expect it to indeed be the case that these matrices are close to diagonal, since modes generally do not couple to each other unless their propagation coefficients are close and thus inter-group mode coupling is the strongest effect inside a MMF; it has also been experimentally observed that for small perturbations, the transmission matrix changes only a small amount [24–26]. However, looking at Eq. (4), there is the further complication that calculating $\mathbf{P_r}$ requires knowing $\mathbf{m}'_{out,d}$; thus it is necessary to only calculate an approximate $\tilde{\mathbf{P}}_\mathbf{r}$ by using $\mathbf{m}_{out,d}$ instead of $\mathbf{m}'_{out,d}$. In other words, we assume no perturbation of the distal modes when calculating the perturbation of the reflection matrix. Unfortunately, this is a less justified assumption than the assumption that $\mathbf{P_r}^T$ is similar to $\mathbf{P_t}$, and we expect this to result in lower-quality spots; we show the effect of this assumption in Section 5, and suggest possible alternatives to improving spot quality in Section 6.

The strategy for compensating for bending of the MMF can thus be summed up as follows: we know the initial configuration **TC** and **RO** using coherent measurements of the unperturbed MMF. When the MMF is perturbed, we would like to know the perturbation $\mathbf{P_t}$ of the transmission matrix in order to perform perfect imaging, but we cannot measure it because we do not have access to the distal end. Instead we use the approximate perturbation $\tilde{\mathbf{P}}_\mathbf{r}^T$ of the reflection matrix, which we can measure because of the partial reflector, as an approximation of $\mathbf{P_t}$ when trying to create spots. With reference to Eq. (3), the spots that we form will then be described by:

$$\tilde{\mathbf{P}}_\mathbf{r} \equiv (\mathbf{M}'_{out,p} \mathbf{M}_{out,d}^{-1})(\mathbf{RO})^{-1}$$
$$\mathbf{m}_{out,d} = \mathbf{P_t T}(\tilde{\mathbf{P}}_\mathbf{r}^T \mathbf{T})^{-1} \tilde{\mathbf{m}}_{out,d} \qquad , \qquad (7)$$

where $\mathbf{M}'_{out,p}$ and $\mathbf{M}_{out,d}$ are matrices formed by many independent measurements of the mode vectors $\mathbf{m}'_{out,p}$ and $\mathbf{m}_{out,d}$, measured using the procedure described in Section 3. This is contrasted with the method of trying to form spots without accounting for the perturbation, as for example in [24], which results in the following:

$$\mathbf{m}_{out,d} = \mathbf{P_t}\mathbf{T}(\mathbf{T}^{-1})\tilde{\mathbf{m}}_{out,d} \quad . \tag{8}$$

Both of these can be contrasted with creating spots if $\mathbf{P_t}$ is exactly known, which results in:

$$\mathbf{m}_{out,d} = \mathbf{P_t}\mathbf{T}(\mathbf{P_t}\mathbf{T})^{-1}\tilde{\mathbf{m}}_{out,d} \quad . \tag{9}$$

Of course, it is not possible to measure $\mathbf{P_t}$ in an endoscopic application. We would expect that Eq. (8) gives the worst results, Eq. (7) gives improved results, and Eq. (9) gives close to ideal results. We experimentally verify these predictions in the following section.

## 5. Experimental results

We affix the MMF to an optical table and displace the center of the MMF laterally by a distance between 0-50 mm, resulting in the approximate change to its shape shown in Fig. 3. For each perturbation, we measure the matrices $\mathbf{TC}$, $\mathbf{RO}$, $\mathbf{P_tTC}$ and $\mathbf{P_rRO}$, which, as mentioned in Section 4, is sufficient to fully characterize the MMF. One set of measurements of these matrices for one such perturbation is shown in Fig. 4.

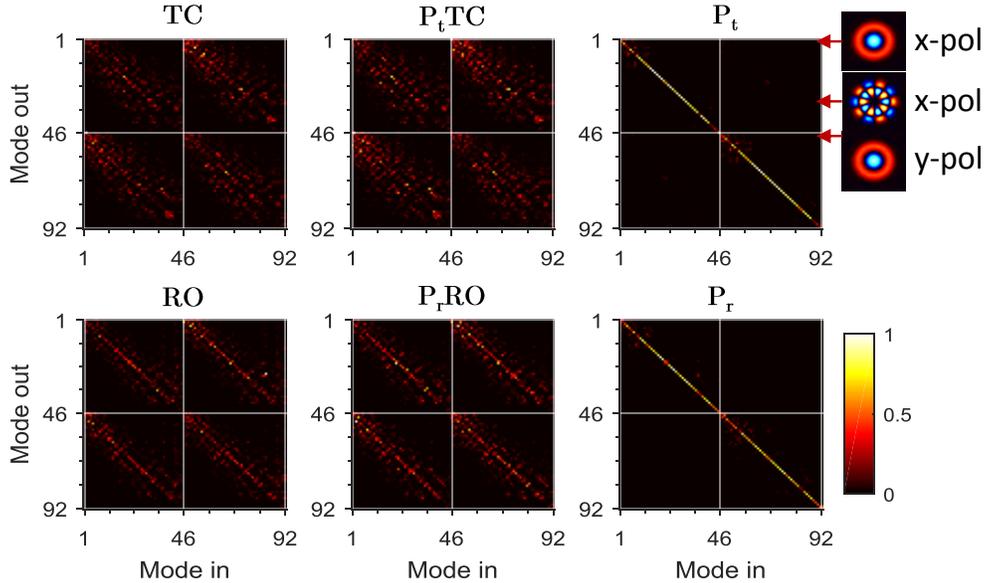

Fig. 4. Transmission (**T**) and reflection (**R**) matrices, perturbed transmission ($\mathbf{P_tT}$) and reflection ($\mathbf{P_rR}$) matrices, and the perturbation matrices for transmission ($\mathbf{P_t}$) and reflection ($\mathbf{P_r}$). Note that we cannot separate **T** or **R** from the input coupling (**C**) and partial reflector (**O**) matrices. Matrix entries correspond to power coupling. Modes 1 to 46 are ordered by decreasing propagation constant in the $x$ polarization; modes 47 to 92 are ordered by decreasing propagation constant in the $y$ polarization. Scale bar represents the energy [0 to 1] in each mode.

As can be seen in Fig. 4, there is substantial mode coupling for both **TC** and **RO**. This could be due to the U-bend in the MMF, micro-bends of the MMF during manufacture, or mismatched coupling into the MMF represented by the matrices **C** and **O**. While we cannot precisely separate these effects, the fact that **TC** is much less diagonal than **RO** would suggest that much of the mode coupling in the matrices **TC** and $\mathbf{P_tTC}$ is due to **C** (mismatched coupling). This hypothesis is supported by the off-diagonal "lines" that can be seen in **TC** and $\mathbf{P_tTC}$, which we would expect to happen due to mismatched coupling between spatial modes that are similar to each other spatially. By contrast, bending the MMF results in perturbation matrices $\mathbf{P_t}$ and $\mathbf{P_r}$ that are strongly diagonal. We also note that most perturbation occurs in the lower-order modes while higher-order modes remain unaffected. The loss that can be seen in the higher-order modes of $\mathbf{P_t}$ and $\mathbf{P_r}$ is due to loss from propagation through the MMF.

We then use $\mathbf{P_r}$ to experimentally create a grid of spots at the distal end of MMF2 (as described at the end of Section 4), which is shown in Fig. 5. In Fig. 5, column 2 shows $\mathbf{P_t T}(\mathbf{P_r^T T})^{-1}$ and three representative spots we create using the matrix. For comparison purposes, column 1 shows the matrix and spots corresponding to no compensation, and column 3 shows the matrix and spots corresponding to perfect compensation, as described by Eq. (8) and Eq. (9) respectively.

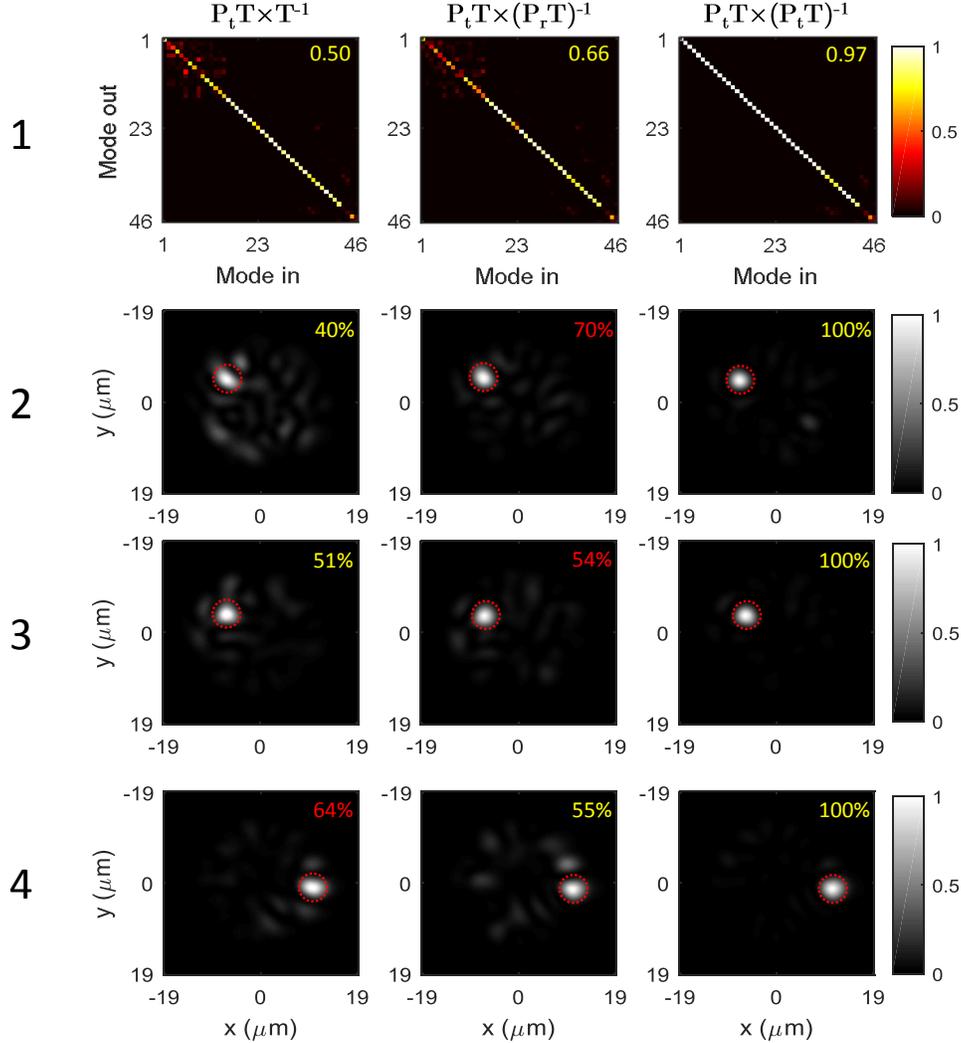

Fig. 5. Row 1: matrices corresponding to no compensation, perturbation compensation, and ideal compensation. The correlation coefficient between each matrix and the identity matrix is indicated. Rows 2 to 4: representative spots we excite in polarization $x$ using the method corresponding to the matrix in the column above. The amount of power contained in the spot (relative to perfect compensation) for each excitation is indicated. Row 2 shows a spot that is substantially better when we use compensation, row 3 shows a spot that is about the same whether we use compensation or not, and row 4 shows a spot that is better without compensation. Averaged across all spots, perturbation compensation results in 30% more power contained inside the spot as compared to no compensation, almost exactly equal to the ratio of the matrix correlation coefficients between compensation and no compensation. The linear polarizer is set to 0° for all spot excitations. Red indicates the method (compensation or no compensation) yielding higher power in the spot.

As predicted, using perturbation compensation results in a substantially better matrix and better spots than no compensation. Both the matrix correlation coefficient metric and average spot power metric exhibit approximately the same improvement, about 30%, as predicted by theory. We normalize the total power contained in a spot for both compensation and no compensation to the percentage of power contained in a spot for perfect compensation. The percentage of total power contained in a spot for perfect compensation is limited to around 50% because, as described in Section 3, only 85% of the SLM is available for the formation of spots (this is not an inherent limitation, but an implementation choice here) and we can only launch one polarization at a time into the MMF; this limits the ensemble of modes that we can excite. The fact that $\mathbf{P_t T} \times (\mathbf{P_t T})^{-1}$ is not equal to the identity matrix (Fig. 5, row 1) indicates that $\mathbf{T}$ has strong loss for its highest-order modes. Our results also agree with previously published results that spots do not completely disappear even for relatively large uncompensated perturbations [24].

Finally, in Fig. 6, we compare the correlation coefficients of the matrices corresponding to no compensation, perturbation compensation using $\mathbf{P_r}$, perturbation compensation using $\tilde{\mathbf{P}}_\mathbf{r}$, and perfect compensation for various perturbations of the MMF. All matrix correlation coefficients are computed element-wise using the Frobenius norm.

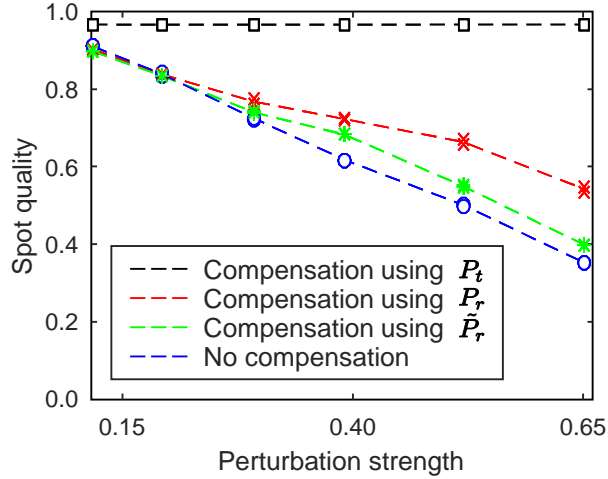

Fig. 6. Comparison of the quality of spot formation for different compensation methods vs. the perturbation strength. Each plot includes two points per perturbation, corresponding to exciting spots in the x- and y-polarizations. The y-axis (spot quality) is parameterized by the correlation coefficient between row 1 of the matrix shown in Fig. 5 and the identity matrix. The x-axis (perturbation strength) is parameterized by the one minus the correlation coefficient between $\mathbf{P_t}$ and the identity matrix. The MMF exhibits non-monotonic behavior, in that larger physical perturbations do not necessarily lead to more substantially different perturbation matrices.

It can be seen that perturbation compensation of the matrix using $\mathbf{P_r}$ yields very substantial improvements of the spot quality for large perturbations and no loss in spot quality for small perturbations. Compensation of the matrix using the approximate reflection matrix $\tilde{\mathbf{P}}_\mathbf{r}$ is worse, as expected, but still shows improvement. For all matrix inversions, we use the singular-value decomposition with up to 90% of the singular values. For $\mathbf{P_r}^{-1}$ and $\tilde{\mathbf{P}}_\mathbf{r}^{-1}$ we take the extra step of averaging the matrix inverse with $\mathbf{I}$; this procedure makes the matrix inverse slightly worse for large perturbations, but more stable for small perturbations, and so ensures that perturbation compensation gives better results for all perturbations.

## 6. Discussion

The experimental results of Section 4 demonstrate substantial improvement of the ability to create spots at the distal end of a perturbed fiber by using information from the reflection matrix, particularly for large perturbations. This may be surprising, given the number of approximations made in Section 3. Further, the worst-case scenario of substantial inter-group mode coupling and fully random, non-diagonal perturbation matrices described in [27] is found to be pessimistic; in the experiments presented here, inter-group mode-coupling is found to be weak, and perturbation mainly affects the coupling between lower-order modes. Thus it is sufficient here to use a simple partial reflector instead of a structured reflector as described in [27]. We note that it is feasible to use either a partial reflector or a structured reflector. For example, a structured reflector can be fabricated by coating chrome on a glass or fused silica cover slip, then using electron-beam lithography and a lift-off standard patterning process to create two-dimensional patterns with 0.5-2 μm feature size and transmission coefficient in the desired range. Such a structured reflector can be easily diced and attached to the end of the MMF [31]. Alternatively, the end of the MMF can be coated with a thin layer of chrome, which can be patterned using a focused ion beam to realize a compact structured reflector.

While the method we describe for perturbation compensation is useful in increasing the SNR for the creation of spots as compared to no compensation, it falls short of a fully realized method to completely compensate for the perturbations of a MMF and so enable fully flexible MMF imaging. Notably, the use of the approximate reflection matrix perturbation $\tilde{\mathbf{P}}_r$ instead of the exact matrix $\mathbf{P}_r$ leads to much reduced improvement in quality of spot formation, which is not unexpected given the major assumptions made, as described in Section 3. However, our results suggest more promising methods to achieve accurate spots beyond simply assuming that the perturbation of the reflection matrix is equal to the perturbation of the transmission matrix. For example, a straightforward method would be to measure, while the endoscope is not in use, the transmission and reflection matrices for a large number of perturbations; then, assuming that the space of perturbations is sampled well enough, when the endoscope is in use, one could interpolate the expected transmission matrix by measuring the reflection matrix. This method only requires that the transmission and reflection matrices map to each other one-to-one, which we would expect to be the case from the small space of perturbations we observe, and that the results are repeatable. Of course, this method is computationally inefficient; it might also be possible to use more complex and robust algorithms to analyze the transmission and reflection matrices and find a mathematical relationship between them, similar to recent work that uses machine learning to transmit images through random media [32,33].

A pressing question is how these results scale up as the number of modes $N$ of the MMF increases. A useful imaging system is likely to require at least $N = 1000$ modes, and ideally would employ many more. As $N$ scales up, the imaging time will increase by $O(N)$, which is acceptable given recent developments in imaging speed mentioned previously. Image processing time will increase by $O(N^2)$, which is acceptable given that the algorithm is highly parallelizable. The key question, then, is how much $\mathbf{P}$ and $\mathbf{T}$ will change for a MMF with many more modes. For example, it is possible that mainly the lowest-order modes will undergo significant perturbation in mode coupling, as we observe here? In that case, we would expect perturbation compensation to work as well as it does here. Conversely, MMFs with more modes could be much more sensitive to perturbations, in which case the method demonstrated here would not work. Extrapolating from recent research, the more optimistic scenario seems more likely; it is possible to substantially increase the NA of the MMF and thus the number of modes $N$ without affecting the strength of mode coupling [24–26].

The results presented here could be improved upon further by reducing experimental error, sources of which include imperfect coupling of light into the MMF, phase jitter (a vibration at a frequency of 11 Hz is observed), drift of the laser power, spherical aberration of the lenses, and technical limitations of the SLM and camera.

## 7. Conclusion

We have proposed and experimentally verified a method to obtain information about the transmission matrix of a perturbed step-index MMF using a partial reflector attached to the end of the MMF. We experimentally find that the perturbation of the reflection matrix is highly correlated with the perturbation of the transmission matrix. Exploiting this observation, we demonstrate a simple method that allows the transmission matrix to be calculated and spots to be created with 10-15% higher quality compared to methods that assume no change in the MMF. Future methods could better exploit this correlation by using more advanced algorithms to better predict the transmission matrix from the reflection matrix, which could enable imaging through fully flexible MMFs.

## Acknowledgements

We would like to thank Iliya Sigal of the University of Toronto for his help with the camera and optical setup.